\documentstyle[psfig]{caosp}

\begin{document}
\htitle{A search for magnetic stars in late stages of stellar evolution}
\hauthor{V.G. Elkin}
\title{A search for magnetic stars in late stages of stellar evolution}
\author{V.G. Elkin}
\institute{Special Astrophysical Observatory, Nizhnij Arkhyz,
Karachaevo-Cherkesia, Russia}

\date{\today}
\maketitle

Two well-known types of stars with strong dipolar magnetic fields
have been extensively studied.
These are the magnetic CP stars and the magnetic white dwarfs.
In a number of papers, it is suggested that these types of stars may be
related by evolution: CP stars could be progenitors of magnetic white dwarfs
(Angel et al. 1981).
Investigation showed that magnetic fields in CP stars do not change
significantly during their lifetime on the main sequence
 (Glagolevskij et al. 1986).

What happened with the magnetic field later on, after the star has left the
main sequence?
St\c{e}pie\'n (1993) proposed that some very slowly rotating yellow giants
 or subgiants with strong
chromospheric and spot activity, could be former
 Ap stars with strong magnetic fields.

When the stars become red giants, they undergo a deep convective mixing
(first dredge-up) at the red giant branch (RGB), before helium ignition in the 
core. At the asymptotic giant branch (AGB), they have a second dredge-up
(Iben 1991). The internal structure of the star significantly changes in this
case.

After the RGB and AGB, the star looses a large part of its own mass.
If the field is not completely destroyed at these episodes,
 it must also change much.
This can be traced if the magnetic fields are investigated in stars
of intermediate evolution stages, between the main sequence and white dwarfs.

In this context, it is our opinion that there are two types of stars
which should be examined for the presence of strong dipolar magnetic
fields: the horizontal branch A and B stars, and the hot
subdwarfs. These stars are appropriate for direct magnetic
measurements using Zeeman spectroscopy, since they show strong and
narrow absorption lines in their spectra. In addition, some of them may
possess chemical abundance anomalies similar to main sequence CP
stars (Bashek and Sargent 1976, Kodaira and Philip 1984, Heber 1992),
 suggesting an intriguing link between the CP stars and the
evolved A and B stars.

Here, we present the results of a
continuing search for strong magnetic fields in horizontal
branch stars and in hot subdwarfs, using Zeeman spectra obtained
at the 6-metre telescope.

\begin{table}[t]
\small
\begin{center}
\caption{ Stellar parameters of Field Horizontal Branch stars }
\label{Privet}
\begin{tabular}{l|r|r|c|c|c|c}
\hline
Star & $M_v$ & $T_{\rm eff}$ & log g  & $vsini$ & [Fe/H] & $M/M_\odot$ \\
\hline
HD 60778  & 0.95 & 8200 & 3.2 & 17 & -0.5  &  \\
HD 74721  & 1.0  & 8700 & 3.6 &  9 & -1.11 & \\
HD 86986  & 0.9 &7900 & 3.1 & 20 & -1.3  & 0.48 \\
HD 97859  &      & 15300& 4.0 & 105& -1.2  &   \\
HD 109995 & 1.1  & 8300 & 3.2 & 15 & -1.3  & 0.37 \\
HD 117880 & 2.5  & 8400 & 3.6 &    & -1.7  &   \\
HD 161817 & 0.8 & 7500 & 2.9 & 10 & -1.0  & 0.42    \\
HD 169027 & 0.3 & 11600 & 3.8  &&                   \\
\hline
\end{tabular}
\end{center}
\end{table}

\begin{table}[t]
\small
\begin{center}
\caption{Magnetic fields and radial velocities of FHB stars}
\bigskip
\begin{tabular}{l|r|r|c}
\hline
Star &  $B_e$ (G)& $\pm \sigma$ (G) & $RV$ (km/s)  \\
\hline
 HD 60778  &      +150 &  115   &  +74   \\
 HD 74721  &      +240 &  150   &  +40   \\
 HD 86986  &      -430 &  580   &  +3    \\
 HD 97859  &      -400 &    &  +66   \\
\hline
 HD 109995 &      -820 &  470   &          \\
	   &      +700 &  520 &  -139  \\
\hline
 HD 117880 &           &      &  +142   \\
\hline
 HD 161817 &      +30  &  110 &            \\
	   &       -550 &  140 &  -368  \\
	   &       -90  &  80    &               \\
	   &       +100 &  160   &  -366  \\
\hline
 HD 169027 &       -320 &  820   &  -26   \\
\hline
\end{tabular}
\end{center}
\end{table}

\begin{table}[t]
\small
\begin{center}
\caption{Parameters for hot subdwarfs}
\begin{tabular}{l|l|c|c|c}
\hline
Star & Sp & $M_v$ & $T_{\rm eff}$ & log g   \\
\hline
HD~4539        & sdB & 3.6  & 25000 & 5.4 \\
Feige 87       & sdB &      & 28000 & 5.6  \\
HD 76431       & sdB & 2.0  & 35000  & 4.5   \\
BD+$75^\circ325$ & sdO & 4.0  & 50000 & 5.3 \\
BD+$25^\circ2534$& sdOp & 4.1 & 34000 & 5.5 \\
BD+$25^\circ4655$& sdO &  & 42200 & 6.7 \\
HD 128220      & sdO+G&  & 42500 & 4.5 \\
HD 149382      & sdOB & 4.5 &  35000 & 5.5  \\
\hline
\end{tabular}
\end{center}
\end{table}

\begin{table}[t]
\small
\begin{center}
\caption{ Magnetic field measurements of hot subdwarfs}

\begin{tabular}{l|r|r|l}
\hline
\noalign{\smallskip}
Star & $B_{\rm e}$ (G)& $ \pm \sigma \rm (G)$ & Comment \\
\noalign{\smallskip}
\hline\noalign{\smallskip}
HD~4539     &  -1300  &  2100  & $\lambda 4000-4600$ \AA \\
	    & +670 & 500  & $HeI~\lambda~5876 $ 2 spectra\\
\hline
Feige 87     &  -1800  &  3500  &  H-magnetometer.     \\
\hline
HD 76431   &   -50   &   130 &   $\lambda 4000-4600$ \AA \\
\hline
$\rm {BD+75^\circ325}$   &   +1260 &   870 & $\lambda 4000-4600$ \AA \\
			 & -1680 &  60 &$HeI~\lambda~5876 $ 3 spectra\\
		  & +970 & 140  &$ HeI~\lambda~4471 $ 3 spectra\\
		  &  -80 & 280  &$HeI~\lambda~5876 $ 2 spectra \\
\hline
$\rm {BD+25^\circ2534}$ &  +460 &   & $  H\alpha$ 1 spectrum     \\
		     & +1750 &  230 & $  H\alpha + HeI~\lambda~5876 $ \\

		   & -1300 &  600 & $HeI~\lambda~5876 $ 2 spectra \\
		       & -1100 & 390  & $HeI~\lambda~5876 $ 2 spectra \\
			      & -100 & 400  & $HeI~\lambda~5876, 6678$      \\
\hline
BD+$25^\circ4655$ &  +240 & 340 & $HeI~\lambda~6678 $ 2 spectra \\
		  &  400  & 280 & $\lambda 4400-4500 $ \AA   \\
		  &  -400 & 240 &  $\lambda 4400-4500 $ \AA  \\
\hline
HD 128220 &   -520  &   950 & $\lambda 4000-4600 $ \AA  \\
	  &   -340  &   400 & $HeI~\lambda~5876 $ 3 spectra  \\
	  &   170   &       & $HeI~\lambda~5876 $ 1 spectrum  \\
\hline
HD 149382 &   -10   &   890 & $\lambda 4000-4600$ \AA \\
	  & -1200   &   900 & $HeI~\lambda~5876 $ 3 spectra \\
\hline
\end{tabular}
\end{center}
\end{table}

{}

\end{document}